\documentclass[11pt,fleqn]{article}
\usepackage{amsmath,amsfonts,amssymb,latexsym}

\usepackage{graphicx}
\usepackage{mathptmx}
\usepackage{bm}
\usepackage{amssymb}
\usepackage{amsmath}

\newcommand {\cG}{\cal G}
\newcommand {\cD}{\cal D}
\newcommand {\bg}{\bar \gamma}
\newcommand {\bp}{\bar \psi}

\def\myfigure #1#2#3#4
{\begin{figure}[ht]\begin{center}
			\includegraphics[width=#2 \textwidth]{#1.jpg}
			\parbox[t]{#4\textwidth}{\caption{#3}\label{#1}}
\end{center}\end{figure}}

\def \myfigures #1#2#3#4#5#6#7#8
{\begin{figure}[ht]
		\begin{center}
			\includegraphics[width=#2 \textwidth]{#1.jpg}
			\hfill
			\includegraphics[width=#6 \textwidth]{#5.jpg}
			\parbox[t]{#4\textwidth}{\caption {#3}\label{#1}}
			\hfill
			\parbox[t]{#8\textwidth}{\caption {#7}\label{#5}}
		\end{center}
\end{figure} }

\begin{document}
	
	\baselineskip -24pt
	\begin{center}
		\title{ Spinor field in Bianchi type-I cosmology with Lyra’s geometry}
		{\bf Bijan Saha}\\
		
		{Laboratory of Information Technologies\\
			Joint Institute for Nuclear Research\\
			141980 Dubna, Moscow region, Russia\\ and\\
			Peoples' Friendship University of Russia (RUDN University)\\
			6 Miklukho-Maklaya Street, Moscow, Russian Federation\\
			orcid: 0000-0003-2812-8930 }\\ 
		email:{bijan@jinr.ru}\\
		URL: {http://spinor.bijansaha.ru}
	\end{center}

	\hskip 1 cm

	\begin{abstract}
		
	In this study, we examine the role of a nonlinear spinor field in the evolution of the Universe within the framework of a Bianchi type-I cosmological model with Lyra’s geometry. Previous research has explored the nonlinear spinor field in various anisotropic and isotropic cosmological models, revealing that the presence of nontrivial, non-diagonal components of the spinor field's energy-momentum tensor imposes severe restrictions on both the space-time geometry and the spinor field itself. In our current study, we find that while these restrictions still apply, the introduction of Lyra’s geometry significantly influences the evolution of the Universe. This influence arises from the fact that the invariants of the spinor field are dependent on the Lyra geometry parameter.
	\end{abstract}
	
	keywords: {spinor field; BI cosmology; Lyra's Geometry}
	
	pacs: {98.80.Cq}
	
	
	\bigskip
	
	\section{Introduction}

The history of science is marked by constant change and refinement. Even Einstein's theory of gravity evolved over time, with him publishing several versions of the theory and making modifications along the way. One such modification was the introduction of the cosmological constant. This evolution reflects the ongoing search for better theories that can deepen our understanding of the natural world. Shortly after Einstein's groundbreaking work, Weyl sought to unify gravity and the electromagnetic field by generalizing Riemannian geometry \cite{Weyl}. However, Weyl's theory was not widely accepted, as it contradicted several well-established observational results.
In 1951, Lyra proposed a modification of Riemannian geometry that closely resembles Weyl's theory \cite{Lyra}. Unlike Weyl geometry, however, Lyra's approach maintains a metric-preserving connection, similar to that of Riemannian geometry. In doing so, he introduced a gauge function into the otherwise structureless manifold. Lyra's theory was further developed by Sen \cite{Sen1957}, Halford \cite{Halford}, Sen and Dunn \cite{Sen1971}, Sen and Vanstone \cite{Sen1972}, and others. In recent years, Lyra's geometry has found extensive applications in cosmology \cite{Beesham,Jahromi,Bakry,Shchigolev}.

In recent years, the spinor field has gained significant attention in cosmology due to its ability to model a variety of source fields, such as perfect fluids and dark energy \cite{Saha1997GRG,Saha1997JMP,SahaPRD2001,Greene,ELKO,PopPRD}. Several studies have shown that the spinor field is highly sensitive to gravitational effects, with the nontrivial, non-diagonal components of its energy-momentum tensor (EMT) imposing severe constraints on both the space-time geometry and the nonlinearity of the spinor field itself \cite{Saha2016,Saha2018}. The spinor field within Lyra's geometry has also been explored in previous works \cite{Casana}. The primary aim of this study is to investigate whether the introduction of Lyra's geometry into the system can alleviate the restrictions typically found in the standard treatment of spinor fields.

\section{Basic equations} 

In this section we give a short description of Lyra's geometry, spinor field and obtain the gravitational field equations for a Bianchi type-I cosmological model. 

\subsection{Lyra's geometry}
Lyra suggested a modification of Riemannian geometry which is also a modification of Weyl geometry. The metrical concept of gauge in Weyl geometry was modified by a structureless gauge function. According to Lyra's geometry the displacement vector from a point $P(x^\mu)$ to a neighbouring point $P'(x^\mu + dx^\mu)$ is defined by $\xi^\mu = x^0 dx^\mu$, where $x^0$ is a nonzero analytical function of coordinates and fixes the gauge of the system. Together with coordinate system $x^\mu$,\, $x^0$ form a so-called reference system $(x^0,x^\mu)$. The transformation to a new reference system is given by 
\begin{align}
	x^\mu = x^\mu(\bar x^1,\cdots ,\bar x^n), \quad 	x^0 = x^0(\bar x^1,\cdots ,\bar x^n,\bar x^0), \label{ref}
\end{align} 
where $\partial x^0/\partial {\bar x^0} \ne 0$ and $\det \partial x^\mu / \partial {\bar x^\nu} \ne 0$. Under the transformation \eqref{ref} a contravariant vector $\xi^\mu$ is transformed according to 
\begin{align}
	\bar \xi^\mu = \lambda A^\mu_\nu \xi^\nu, \quad A^\mu_\nu = \frac{\partial \bar x^\mu}{\partial x^\nu}, \quad \lambda = \frac{\bar x^0}{x^0}, \label{trans0}
\end{align}
with $\lambda$ being the gauge factor of transformation. 

In any general reference system $(x^0,\, x^\mu)$ the infinitesimal parallel transfer of a vector from  $P(x^\mu)$ to $P'(x^\mu + dx^\mu)$ can be expressed as
\begin{align}
\delta \xi^\mu = -\tilde\Gamma^\mu_{\alpha \beta} \xi^\alpha x^0 dx^\beta, \quad 
	\tilde\Gamma^\mu_{\alpha \beta} &= \Gamma^\mu_{\alpha \beta} - \frac{1}{2} \delta^\mu_\alpha \phi_\beta, \quad \phi_\alpha =  - \frac{\partial (\log \lambda^2)}{\partial x^\alpha}. \label{conL} 
\end{align}	
It should be noted that $\Gamma^\mu_{\alpha \beta} = \Gamma^\mu_{\beta \alpha}$, though $\tilde \Gamma^\mu_{\alpha \beta} \ne \tilde \Gamma^\mu_{\beta \alpha}.$ 

Since the displacement vector between two neighbouring points $P(x^\mu)$ and $P'(x^\mu + dx^\mu)$ in this case is define by $\xi^\mu = x^0 dx^\mu$, the interval between them is given by the invariant
\begin{align}
ds^2 = g_{\mu \nu} x^0 dx^\mu x^0 dx^\nu, \label{met} 
\end{align}
where $g_{\mu\nu}$ is the symmetric tensor of second rank. 
The parallel transport of length in Lyra geometry is integrable, i.e.,
$\delta(g_{\mu\nu} \xi^\mu \xi^\nu) = 0$ and the connection $\Gamma^\alpha_{\mu\nu}$ in  \eqref{conL} takes form		

\begin{align}
\Gamma^\alpha_{\mu\nu}  = \frac{1}{x^0} \{^\alpha_{\mu\nu}\} + \frac{1}{2}\left(\delta^\alpha_\mu \phi_\nu + \delta^\alpha_\nu \phi_\mu - g_{\mu\nu} \phi^\alpha\right),  \label{connecLyr} 
\end{align}
which is similar to that of Weyl geometry except the multiplier $1/x^0$. Here $\{^\mu_{\alpha \beta}\}$ is the Levi-Civita connection.  Note that in Lyra geometry the derivative $\partial/\partial x^\mu$ is substitute by $\partial/(x^0 \partial x^\mu)=(1/x^0) \partial/ \partial x^\mu $.	

The parallel transfer, hence the equation of motion 
\begin{align}
\frac{1}{x^0} \frac{\partial \xi^\alpha}{\partial \beta} + \tilde \Gamma^\alpha_{\nu \beta} \xi^\nu = 0, \label{em} 
\end{align} 
can be integrated if the components of the tensor 
\begin{align}
	K^\lambda_{\mu\alpha\beta} &= \frac{1}{(x^0)^2} \left[\frac{\partial(x^0 \tilde \Gamma^\lambda_{\mu\beta})}{\partial x^\alpha} - \frac{\partial(x^0 \tilde \Gamma^\lambda_{\mu\alpha})}{\partial x^\beta} + x^0 \tilde \Gamma^\lambda_{\rho\alpha}
	x^0 \tilde \Gamma^\rho_{\mu\beta} - x^0 \tilde \Gamma^\lambda_{\rho\beta} x^0 \tilde \Gamma^\rho_{\mu\alpha}\right], \label{curvature} 
\end{align} 
vanish \cite{Sen1957}. Einstein's field equation in Lyra's geometry in normal gauge ($x^0 = 1$) was found by Sen \cite{Sen1957} and can be written as 

\begin{align}
G_\mu^\nu + \frac{3}{2} \phi_\mu \phi^\nu - \frac{3}{4}  
\delta_\mu^\nu \phi_\alpha \phi^\alpha  = \kappa
T_\mu^\nu, \quad G_\mu^\nu =	R_\mu^\nu - \frac{1}{2} \delta_\mu^\nu R. \label{EE}
\end{align}
where $\phi_\mu$  is the displacement vector.

\subsection{ Spinor field}

We consider the spinor field Lagrangian given by \cite{SahaPRD2001}
\begin{align}
	L_{\rm sp} = \frac{\imath}{2} \biggl[\bp \gamma^{\mu} \nabla_{\mu}
	\psi- \nabla_{\mu} \bar \psi \gamma^{\mu} \psi \biggr] - m_{\rm sp}
	\bp \psi - \lambda F, \label{lspin}
\end{align}

 where the nonlinear term $F$ describes the self-interaction of a spinor field and can be
	presented as some arbitrary functions of invariants
$K$ that takes one of the following values $\{I,\,J,\,I+J,\,I-J\}$ generated from the real
	bilinear forms of a spinor field. We also consider the case
	$\psi = \psi(t)$ so that $I = S^2 =
		(\bp \psi)^2,\, \&\,\, J = P^2 = (\imath \bp \gamma^5
		\psi)^2$. Here $\lambda$ is the self-coupling constant.

The spinor field equations take the form 
	\begin{align}
	\imath\gamma^\mu \nabla_\mu \psi - m_{\rm sp} \psi - {\cD} \psi -
	\imath {\cG} \gamma^5 \psi & = 0, \label{speq1} \\
	\imath \nabla_\mu \bp \gamma^\mu +  m_{\rm sp} \bp + {\cD}\bp +
	\imath {\cG} \bp \gamma^5 & = 0. \label{speq2}
	\end{align}

where we denote ${\cD} = 2 S F_K K_I$ and ${\cG} = 2 P F_K 	K_J$ with  $F_K = dF/dK$, $K_I = dK/dI$ and
$K_J = dK/dJ.$ In the Lagrangian \eqref{lspin} and spinor field equations \eqref{speq1} and \eqref{speq2}
$\nabla_\mu$  is the covariant covariant derivative of the spinor field so that $\nabla_\mu \psi =
	\partial_\mu - \Omega_\mu \psi$ and $\nabla_\mu \bp = 	\partial \bp + \bp \Omega_\mu$. 
In Lyra's geometry we should substitute $\Gamma^{\rho}_{\mu\nu}$ with $\tilde \Gamma^{\rho}_{\mu\nu} = 
\Gamma^{\rho}_{\mu\nu} - (1/2)\delta^\rho_\mu \phi_\nu =
\frac{1}{x^0} \{^\alpha_{\mu\nu}\} + \frac{1}{2}\left(\delta^\rho_\nu \phi_\mu - g_{\mu\nu} \phi^\rho\right)$.  Thus in this case we have 	
\begin{align}
\tilde\Omega_\mu &= \frac{1}{4} \bg_{a} \gamma^\nu \partial_\mu e^{(a)}_\nu
- \frac{1}{4} \gamma_\rho \gamma^\nu \tilde \Gamma^{\rho}_{\mu\nu} \nonumber\\
& =  \frac{1}{4} \bg_{a} \gamma^\nu \partial_\mu e^{(a)}_\nu
- \frac{1}{4} \gamma_\rho \gamma^\nu \{^{\rho}_{\mu\nu}\} + \frac{1}{8}\left( \gamma_\rho \gamma^\rho \phi_\mu -\gamma_\rho \gamma_\mu \phi^\rho\right).
\label{sfc}
\end{align}

The energy momentum tensor of the spinor
field is given by 

\begin{align}
T_{\mu}^{\,\,\,\rho} & = \frac{\imath g^{\rho\nu}}{4}  \left(\bp
\gamma_\mu \nabla_\nu \psi + \bp \gamma_\nu \nabla_\mu \psi -
\nabla_\mu \bar \psi \gamma_\nu \psi - \nabla_\nu \bp \gamma_\mu
\psi \right) \,- \delta_{\mu}^{\rho} L_{\rm sp} \nonumber\\
& = \frac{\imath}{4} g^{\rho\nu} \left(\bp \gamma_\mu
\partial_\nu \psi + \bp \gamma_\nu \partial_\mu \psi -
\partial_\mu \bar \psi \gamma_\nu \psi - \partial_\nu \bp \gamma_\mu
\psi \right)\nonumber\\
& -  \frac{\imath}{4} g^{\rho\nu} \bp \left(\gamma_\mu
	\tilde\Omega_\nu + \tilde\Omega_\nu \gamma_\mu + \gamma_\nu \tilde\Omega_\mu +
	\tilde\Omega_\mu \gamma_\nu\right)\psi
\nonumber\\
& -  \delta_{\mu}^{\rho} \lambda \left( 2 K F_K - F(K)\right). \label{temsp0}
\end{align} 

On account of spinor field equations \eqref{speq1} and \eqref{speq2} the spinor
field Lagrangian takes the form $L_{\rm sp} =  2 K
F_K - F(K).$ Thanks to spinor field equations the
conservation of energy holds, i.e.,
\begin{align} 
	T^{\mu}_{\nu; \mu} = 0. \label{conserv}
\end{align}
As we see later, this very fact will help us to define the parameter of Lyra's geometry in terms of volume scale. 

\subsection{Bianchi type-I model}

The Bianchi type-I we take in the form

\begin{align}
	ds^2 = dt^2 - a_1^2 dx_1^2 - a_2^2 dx_2^2 -
	a_3^2 dx_3^2, \label{bi}
\end{align}
 with $a_1,\,a_2$,\, $a_3$  being the functions of time only. 
 Following Sen we consider the gauge function as follows:
\begin{align}
\phi_\mu = \{\beta(t),\,0,\,0,\,0\}, \label{phi}
\end{align}  
 
 The spinor affine connection in this case has the form:
\begin{subequations} 
	\label{sacbi}
	\begin{align}
		\Omega_0 &= -\frac{3}{8} \beta,\label{sac0}\\ 
		\Omega_1 &= \frac{1}{2}\left(\dot a_1 - \frac{\beta a_1}{4}\right) 
		\bg^1\bg^0, \label{sac1}\\
		\Omega_2 &= \frac{1}{2}\left(\dot a_2 - \frac{\beta a_2}{4}\right) \bg^2\bg^0, \label{sac2}\\ 
		\Omega_3 &= \frac{1}{2} \left(\dot a_3 - \frac{\beta a_3}{4}\right) \bg^3 \bg^0. \label{sac3}
	\end{align}
\end{subequations}
Thus we see that the introduction of Lyra geometry bring changes in $\Omega_\mu$. The spinor field equations now take the form 
\begin{subequations} 
	\label{spinbi01}
\begin{align}
\frac{1}{x^0}\dot\psi + \frac{\dot V}{2 V} \psi + \frac{3 \beta}{4} \psi + \imath  \left( m_{\rm sp} + {\cD}\right) \bg^0 \psi - {\cG} \bg^0 \bg^5 \psi & = 0, \label{speq1n} \\
\frac{1}{x^0}\dot\bp + \frac{\dot V}{2 V} \bp  - \imath \left( m_{\rm sp} + {\cD}\right)\bp \bg^0 - {\cG} \bp \bg^0 \bg^5 & = 0. \label{speq2n}
\end{align}
\end{subequations}
Note that in Lyra's geometry the differential operator $\partial/\partial x^\mu$ is substituted by   $ (1/x^0) \partial/\partial x^\mu$. But in natural gauge with $x^0 = 1$ we omit it in our further calculations.

For the invariants we find 

\begin{subequations} 
	\label{invbi01}
	\begin{align}
		\dot S_0 + \frac{3}{4} \beta S_0 + 2 {\cG} A_0^0 & = 0, \label{invS} \\
		\dot P_0 + \frac{3}{4} \beta P_0 + 2 \left( m_{\rm sp} + {\cD}\right) A_0^0 & = 0, \label{invP}\\
		 \dot A_0^0 + \frac{3}{4} \beta A_0^0 + 2 \left( m_{\rm sp} + {\cD}\right) P_0 - 2 {\cG} S_0 & = 0, \label{invA0}
	\end{align}
\end{subequations}
with the solutions 

\begin{align}
	S_0^2  - P_0^2 + {A_0^0}^2 = C_0 \exp[-(3/2) \int \beta(t) dt], \label{invf}
\end{align}  
with $C_0$ being some constant of integration. Here $S_0 = S V$, $P_0 = P V$ and $A_0^0 = A^0 V$, with $V$ being the volume scale:

\begin{align}
	V = a_1 a_2 a_3. \label{Volumescale}
	\end{align}

On account of \eqref{sacbi} we find the following diagonal components of the energy momentum tensor of the spinor field.  

\begin{align}
	T_0^0 = m_{\rm sp} S + \lambda F = \varepsilon, \quad T_1^1 = T_2^2 = T_3^3 =
	\lambda \left(F - 2 K F_K\right) = -p. \label{emtCar}
\end{align}

Note that they are identical to those found in \cite{Saha2016,Saha2018} where Lyra's geometry was not exploited. The reason is the additional terms occurred in this case canceled each other. Same thing happens for non-diagonal components as well. The nontrivial non-diagonal components of EMT in this case read \cite{Saha2016,Saha2018}:

\begin{subequations} 
	\label{emtnd}
\begin{align}
	T_2^1 &=  \frac{1}{4}\frac{a_2}{a_1} \left(\frac{\dot
		a_1}{a_1} - \frac{\dot a_2}{a_2}\right) A^3, \label{emt12} \\ 
	T_1^3 &= \frac{1}{4}\frac{a_1}{a_3}
	\left(\frac{\dot a_3}{a_3} - \frac{\dot a_1}{a_1}\right) A^2, \label{emt31}\\
	T_3^2 &= \frac{}{4}\frac{a_3}{a_2}
	\left(\frac{\dot a_2}{a_2} - \frac{\dot a_3}{a_3}\right)A^1.
	\label{emt23}
\end{align}
\end{subequations}
As we will find later, these non-diagonal components of EMT will characterize the geometry of space-time or spinor field.

\subsection{Einstein equation and its solution}
	
Bianchi type-I space-time \eqref{bi} possesses only diagonal components of Einstein tensor. 
The diagonal components of Einstein's equations are \cite{Saha2016,Saha2018}:

	\begin{subequations}
		\label{EE0}
		\begin{align}
			\frac{\ddot a_2}{a_2} + \frac{\ddot a_3}{a_3} +
			\frac{\dot a_2}{a_2}\frac{\dot a_3}{a_3} - \frac{3}{2} \beta^2 &= \kappa \lambda\left(F - 2 K F_K\right), \label{EE11}\\
			\frac{\ddot a_3}{a_3} + \frac{\ddot a_1}{a_1} +
			\frac{\dot a_3}{a_3}\frac{\dot a_1}{a_1}  - \frac{3}{2} \beta^2 &= \kappa \lambda \left(F - 2 K F_K\right), \label{EE22}\\
			\frac{\ddot a_1}{a_1} + \frac{\ddot a_2}{a_2} +
			\frac{\dot a_1}{a_1}\frac{\dot a_2}{a_2}  - \frac{3}{2} \beta^2 &= \kappa
			\lambda \left(F - 2 K F_K\right), \label{EE33}\\
			\frac{\dot a_1}{a_1}\frac{\dot a_2}{a_2} + \frac{\dot
				a_2}{a_2}\frac{\dot a_3}{a_3} + \frac{\dot a_3}{a_3}\frac{\dot
				a_1}{a_1} + \frac{3}{2} \beta^2 &=   \kappa\left(m_{\rm
				sp} S + \lambda F\right). \label{EE00}
		\end{align}
	\end{subequations}
	Thanks to the fact that $T_1^1 = T_2^2 = T_3^3$ from \eqref{EE11} - \eqref{EE33} 
for metric functions $a_i$ one finds 

\begin{align}
a_i &= X_i V^{1/3} \exp[Y_i \int (1/V) dt], \quad \prod_{i=1}^3 X_i = 1, \quad \sum_{i=1}^3 Y_i = 0 . \label{metf}
\end{align}
Here $X_i$ and $Y_i$ are the constants of integration. 
Thus we see that the metric functions are given in terms of $V$. Hence we have to find the volume scale as well. From \eqref{EE0} for volume scale we obtain 

	\begin{align}
		\ddot V &= \kappa \left[m_{\rm sp} S + 6 \lambda\left(F - K F_K \right) \right] V.  \label{Vfind}
	\end{align}
	As one sees, the equation \eqref{Vfind} does not contain $\beta$, hence identical to the case without Lyra's geometry. 
From \eqref{invbi01} it can be shown that 

			\begin{align}
				S &= \frac{C_0}{V}{\exp{\left[-\frac{3}{4}\int \beta (t) dt\right]}}, \quad 
				K = \frac{C_0^2}{V}{\exp{\left[-\frac{3}{2}\int \beta(t) dt\right]}}, \quad C_0 = {\rm const.} \label{K}
			\end{align}

 Note that for $K  = \{J,\,I\pm J\}$ relations \eqref{K} holds for massless spinor field only, whereas for $K  = I$ it is true for both massive and massless spinor field. Thus we see that the equation \eqref{Vfind} implicitly depends on $\beta$ that defines Lyra geometry.  

Taking into account that the in case of spinor field $T^\nu_{\mu;\nu} = 0$ on accout on Bianchi identity $G^\nu_{\mu;\nu} = 0$ from \eqref{EE} we find

\begin{align}
	\left(\frac{3}{2} \phi_\mu \phi^\nu - \frac{3}{4}  
	\delta_\mu^\nu \phi_\alpha \phi^\alpha\right)_{;\nu}  = V \dot \beta + \beta \dot V = 0, \label{BI}
\end{align}
with the solution

\begin{align}
	\beta &= \beta_0/V, \quad \beta_0 = {\rm const.} \label{beta}
\end{align}
The R.H.S. of equation \eqref{Vfind} depends on V only as now we have 

	\begin{align}
	S &= \frac{C_0}{V}{\exp{\left[-\frac{3\beta_0}{4}\int \frac{dt}{V(t)}\right]}}, \quad 
	K = \frac{C_0^2}{V^2}{\exp{\left[-\frac{3\beta_0}{2}\int  \frac{dt}{V(t)}\right]}}, \quad C_0 = {\rm const.} \label{VK}
\end{align}

In what follows, stead of \eqref{Vfind} we solve the system \eqref{EE0}, numerically.

\subsection{Numerical solutions}

In this section we solve the system \eqref{EE0}. Introducing directional Hubble parameters we rewrite the equations \eqref{EE11}, \eqref{EE22} and \eqref{EE33} as: 

\begin{subequations}
	\label{Num1}
	\begin{align}
		\dot a_1 &= H_1 a_1, \label{H13}\\
		\dot a_2 &= H_2 a_2, \label{H23}\\
		\dot a_3 &= H_3 a_3, \label{H33}\\
		\dot H_1 &= \frac{\kappa}{2} \lambda \left(F - 2 K F_K\right) +\frac{3}{4}\frac{\beta_0^2}{a_1^2 a_2^2 a_3^2} - \frac{1}{2}\left[2 H_1^2 + H_1
		H_2 + H_3 H_1 - H_2 H_3\right],\label{11biH3}\\
		\dot H_2 &= \frac{\kappa}{2} \lambda \left(F - 2 K F_K\right) +\frac{3}{4}\frac{\beta_0^2}{a_1^2 a_2^2 a_3^2} - \frac{1}{2}\left[2 H_2^2 + H_1
		H_2 + H_2 H_3 - H_3 H_1\right],\label{22biH3}\\
		\dot H_3 &= \frac{\kappa}{2} \lambda \left(F - 2 K F_K\right) +\frac{3}{4}\frac{\beta_0^2}{a_1^2 a_2^2 a_3^2} - \frac{1}{2}\left[2 H_3^2 + H_3 H_1 + H_2 H_3 - H_1 H_2 \right].\label{33biH3}
	\end{align}
\end{subequations}
whereas the equation \eqref{EE00} we exploit to find initial value.

To solve the  equation \eqref{Vfind} numerically we have to give the concrete form of spinor field nonlinearity. It was found earlier that the spinor field nonlinearity can simulate different types of source field such as quintessence, Chapligyn gas, modified quintessence,  modified chapligyn gas etc. \cite{Saha2018}:  

\begin{subequations}
\label{DESim}
\begin{align}
F(K) &= \lambda_1 K^{(1+W)/2} - m_{\rm sp} S, \quad W = {\rm const.}  {\rm -\,\, quintessence},  \label{quint} \\
F(K) &= \left(A + \lambda_1 K^{(1+\alpha)/2}\right)^{1/(1+\alpha)}, \,\,\, A >0, \quad 0 \le \alpha \le 1, \,\, 
 {\rm -\,\, Chapligyn\,\,\, gas}, \label{chap0}\\
F(K) &= \lambda_1 K^{(1+W)/2} + \frac{W}{1 + W} \varepsilon_{\rm cr}, {\rm -\,\, modified\,\, quintessence,} \label{modquint}\\
F(K) &=  \left[\frac{B}{1+W} + \lambda_1
K^{(1+\alpha)(1+W)/2}\right]^{1/(1+\alpha)}, \quad 
{\rm -\,\, modified\,\,\, Chapligyn\,\,\, gas}, \label{mchapsp}
\end{align}
\end{subequations}
where $\lambda_1$ is the integration constant.
It should be emphasized that only in case of quintessence the spinor field can me massive, whereas in other cases it is massless. Moreover, the massive term in \eqref{quint} cancels the massive term in Lagrangian 
\eqref{lspin}. Hence we can dully set $m_{\rm sp} = 0$ in our further calculations. Now the equation \eqref{EE00} in view of \eqref{H13}, \eqref{H23},, \eqref{H33} and \eqref{VK} takes the form 

\begin{align}
	H_1 H_2 + H_2 H_3 + H_3 H_1 &= \kappa F - \frac{3\beta_0^2}{2V^2}. \label{H00}
\end{align}
Given the initial values of $a_1,\,a_2,\,a_3,\,H_2,\,H_3$ one can find 

\begin{align}
	H_1 &= \frac{\kappa F - 3\beta_0^2/(2V^2) - H_2 H_3}{H_2 + H_3}, \quad V = a_1 a_2 a_3. \label{H1in}
\end{align}

\begin{figure}
	\centering
	\includegraphics[width=7 cm]{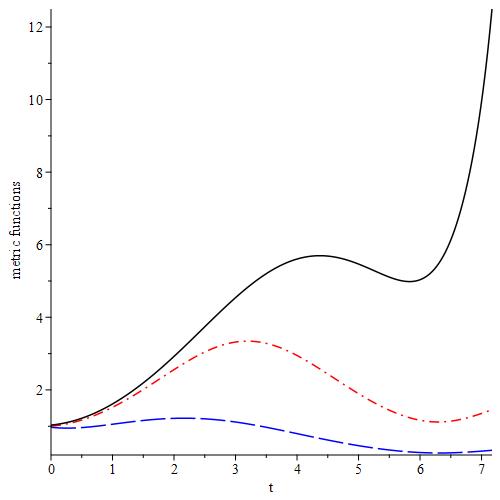}
	\caption{\label{Fig1}Evolution of metric functions $a_1(t)$ (blue long dash),  $a_2(t)$ (red dash-dot) and $a_3(t)$ (black solid) with Lyra geometry when spinor field nonlinearity simulates  modified quintessence.}
\end{figure}

\begin{figure}
	\centering
	\includegraphics[width=7 cm]{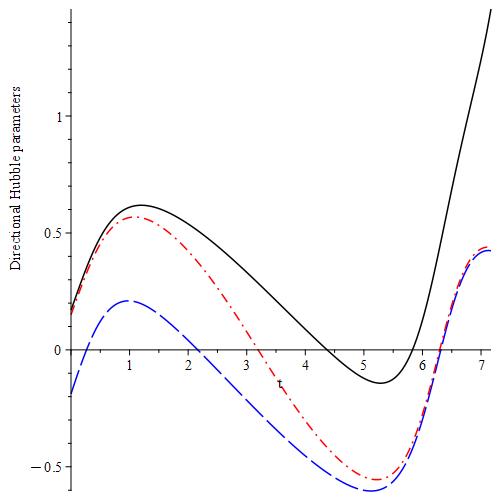}
	\caption{\label{Fig2}Evolution of directional Hubble parameters $H_1(t)$ (blue long dash),  $H_2(t)$ (red dash-dot) and $H_3(t)$ (black solid) with Lyra geometry when spinor field nonlinearity simulates  modified quintessence.}
\end{figure}

In what follows we solve the equation \eqref{Vfind} numerically. Our aim is to clarify whether the introduction of Lyra's geometry bring any essential changes in the solution. For simplicity we set $m_{\rm sp} = 0$, $\kappa = 1$, $\lambda = 1$, $\lambda_1 = 1$. Here we consider modified Chapligyn gas with $B = 1$, $W = 1/2$ and $\alpha = 1/3$. For initial values we set $a_1(0) = 0.97$, $a_2(0) = 1.00$, $a_3(0) = 1.03$, $H_2(0)= 0.15$ and $ H_3(0) = 0.17$. Initial value for $H_1(0)$ is calculated form \eqref{H1in} that in this case gives $H_1(0)= - 1.65$. In figures \ref{Fig1} and \ref{Fig2} we display the evolution of metric functions $a_i(t)$ and directional Hubble parameters $H_i(t)$ for modified quintessence. In the figures blue long das, red dash-dot and black solid line stand components of metric functions and directional Hubble parameters along $X-$, $Y-$ and $Z-$ axes, respectively.

	Let us recall that the spinor field possesses non-diagonal components of EMT which do not depend on spinor mass or spinor field nonlinearity. Given the fact that Einstein tensor for BI model possesses only diagonal components, from \eqref{emtnd} we find 
	
\begin{subequations}	
	\label{NDEE}
\begin{align}
 \left(\frac{\dot a_1}{a_1} - \frac{\dot a_2}{a_2}\right) A^3 &= 0, \label{12} \\
	\left(\frac{\dot a_3}{a_3} - \frac{\dot a_1}{a_1}\right) A^2 &= 0, \label{31} \\
	\left(\frac{\dot a_2}{a_2} - \frac{\dot a_3}{a_3}\right)A^1 &= 0. \label{23}
\end{align}
\end{subequations}
\noindent 	
The foregoing system leads to three different cases:\\
(i) $A^1 = A^2 = A^3 = 0$. By virtue of Fierz identity in this case the spinor field becomes linear and massless;\\
(ii) $A^2 = A^3 = 0$ and $a_2 = a_3$ which gives rise to locally rotational symmetric Bianchi type-I (LRSBI) model;\\
(iii) $a_1 = a_2 = a_3$ i.e. the anisotropy vanishes and BI spacetime becomes Friedmann-Lamaitre-Robertson-Walker (FLRW) one. 

We consider the cases (ii) and (iii) in some forthcoming papers. Here we study the case (i) in details. 
Setting $\lambda = 0$ in \eqref{Num1} we find the following system: 

\begin{subequations}
	\label{Num2}
	\begin{align}
		\dot a_1 &= H_1 a_1, \label{H13l}\\
		\dot a_2 &= H_2 a_2, \label{H23l}\\
		\dot a_3 &= H_3 a_3, \label{H33l}\\
		\dot H_1 &= \frac{3}{4}\frac{\beta_0^2}{a_1^2 a_2^2 a_3^2} - \frac{1}{2}\left[2 H_1^2 + H_1
		H_2 + H_3 H_1 - H_2 H_3\right],\label{11biH3l}\\
		\dot H_2 &= \frac{3}{4}\frac{\beta_0^2}{a_1^2 a_2^2 a_3^2} - \frac{1}{2}\left[2 H_2^2 + H_1
		H_2 + H_2 H_3 - H_3 H_1\right],\label{22biH3l}\\
		\dot H_3 &= \frac{3}{4}\frac{\beta_0^2}{a_1^2 a_2^2 a_3^2} - \frac{1}{2}\left[2 H_3^2 + H_3 H_1 + H_2 H_3 - H_1 H_2 \right].\label{33biH3l}
	\end{align}
\end{subequations}

\begin{figure}
	\centering
	\includegraphics[width=7 cm]{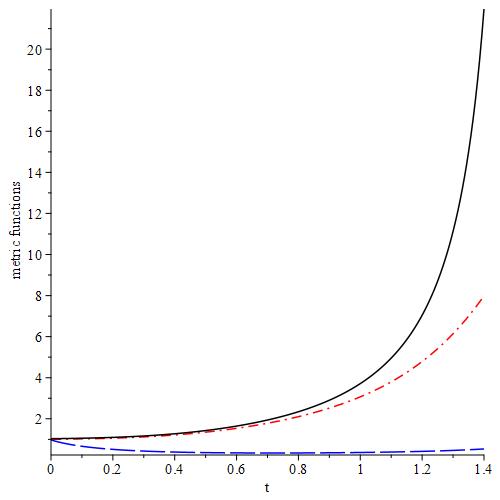}
	\caption{\label{Fig3}Evolution of metric functions $a_1(t)$ (blue long dash),  $a_2(t)$ (red dash-dot) and $a_3(t)$ (black solid) with Lyra geometry in absence of spinor mass and spinor field nonlinearity.}
\end{figure}

\begin{figure}
	\centering
	\includegraphics[width=7 cm]{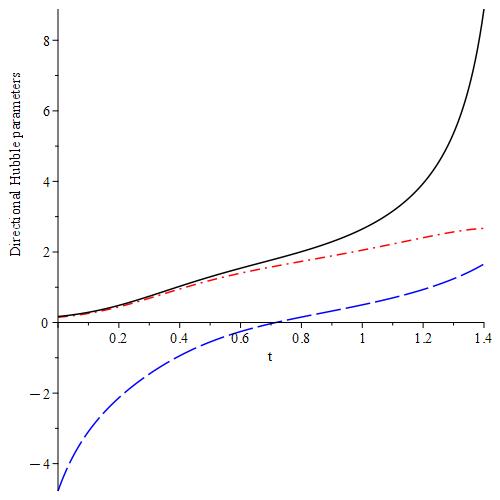}
	\caption{\label{Fig4}Evolution of directional Hubble parameters $H_1(t)$ (blue long dash),  $H_2(t)$ (red dash-dot) and $H_3(t)$ (black solid) with Lyra geometry in absence of spinor mass and spinor field nonlinearity.}
\end{figure}

 As in previous case we solve the system numerically. 
 For initial values we set $a_1(0) = 0.97$, $a_2(0) = 1.00$, $a_3(0) = 1.03$, $H_2(0)= 0.15$ and $ H_3(0) = 0.17$. Initial value for $H_1(0)$ is calculated form \eqref{H1in} that in this case gives $H_1(0)= - 4.77$. In figures \ref{Fig3} and \ref{Fig4} we display the evolution of metric functions $a_i(t)$ and directional Hubble parameters $H_i(t)$ for modified quintessence. In the figures blue long das, red dash-dot and black solid line stand components of metric functions and directional Hubble parameters along $X-$, $Y-$ and $Z-$ axes, respectively.

Note that setting $\beta_0 = 0$ in \eqref{Num1} one can find the system without Lyra's geometry. 

\section{Concluding remarks}

Within the framework of a Bianchi type-I cosmological model, we investigate the role of Lyra's geometry in the evolution of the Universe when it is filled with dark energy modeled by a spinor field. Our analysis reveals that the corresponding Einstein equations retain the same form as in the absence of Lyra’s geometry. However, the dependence of spinor field invariants on the Lyra geometry parameter influences the final results. The relevant equations are solved numerically, with the solutions presented graphically.

As in the standard case, the presence of nontrivial, non-diagonal components in the energy-momentum tensor (EMT) of the spinor field leads to three possible scenarios. In the case of a general Bianchi type-I model, the spinor field becomes massless and linear, a scenario that is also studied numerically. The remaining two cases — the locally rotationally symmetric (LRS) Bianchi type-I model and the Friedmann-Robertson-Walker (FRW) model—will be explored in upcoming papers.

 \vskip 1 cm

\noindent {\bf Funding:} Not applicable

\vskip .5 cm

\noindent {\bf Institutional Review Board Statement:} Not applicable

\vskip .5 cm

\noindent {\bf Informed Consent Statement:} Not applicable

\vskip .5 cm

\noindent {\bf Acknowledgment:}\,\,\,\,{This paper has been
	supported by the RUDN University Strategic Academic Leadership
	Program.}

\vskip .5 cm

\noindent \textbf{DAS:} No datasets were generated or analyzed
during the current study

\vskip .5 cm

\noindent {\bf Conflicts of Interest:} No conflict of interests


\begin{thebibliography}{9999}
	\bibitem{Weyl} H. Weyl, {\it Gravitation and Electricity}, Preuss. Akad. Wiss. Berlin, 465 (1918) 	
	
	\bibitem{Lyra} G. Lyra,  Math. Z. {\bf 54}, 52 (1951)
	
	
		\bibitem{Sen1957} D. K. Sen, Z. Physik. {\bf 149}, 311 (1957)
		
					
				\bibitem{Halford} Halford J. Math. Phys. {\bf 13}, 1699 (1972)
				
					\bibitem{Sen1971} D.K. Sen and K.A. Dunn, J. Math. Phys. {\bf 12}, 578 (1971) 
					
					
		\bibitem{Sen1972} D.K. Sen and J.R. Vanstone, J. Math. Phys. {\bf 13}, 990 (1972)
		
		\bibitem{Beesham} A. Beesham, Aust. J. Phys. {\bf 41}, 833 (1988)  
		
		\bibitem{Jahromi} A.S. Jahromi and H. Moradpour, Int. J. Mod. Phys. D {\bf 27} 1850024 (2018)  
		
			\bibitem{Bakry} M.A. Bakry, Astrophys. Space Sci. {\bf 367}, 35 (2022)
			
				\bibitem{Shchigolev} V.K. Shchigolev and D.N. Bezbatko, Grav. $\&$ Cosmology, {\bf 24}(2), 161 (2018)

\bibitem{Saha1997GRG} Saha B. and Shikin G.N.  Gen. Relat. Gravit. {\bf 29}(9), 1099 (1997)

\bibitem{Saha1997JMP} Saha B. and Shikin G.N. J. Math. Phys. {\bf 38}(10), 5305 (1997)
		
			\bibitem{SahaPRD2001} B. Saha, Phys. Rev. D {\bf 64}, 123501 (2001)
				
		\bibitem{Greene}  Armend$\acute a$riz-Pic$\acute o$n C., Greene P.B.
		Gen. Relat. Grav. {\bf  35}(9), 1637 (2003).
		
		\bibitem{ELKO}  Fabbri L. Phys. Rev. D.  {\bf 85}, 047502 (2012).
		
		\bibitem{PopPRD}  Pop{\l}awski N.J. Phys. Rev. D. {\bf  85}, 107502 (2012).
		
			
			\bibitem{Saha2016} B. Saha, Eur. Phys. J. Plus {\bf 131} 170 (2016)
			
			\bibitem{Saha2018} B. Saha, Phys. Part. Nucl. {\bf 49}(2), 146 (2018)
			
			\bibitem{Casana}  R. Casana, C. A. M. de Melo, B. M. Pimentel, Astrophys. Space Sci. {\bf 305}, 125 (2006)
			

			
	\end{thebibliography}
	\end{document}